\documentclass[11pt, a4paper]{article}%
\usepackage{geometry}
\usepackage{graphicx}
\usepackage{booktabs}
\usepackage{array}
\usepackage{sectsty}
\allsectionsfont{\sffamily\mdseries\upshape} 
\usepackage{amssymb}
\usepackage{amsthm}
\usepackage{geometry}                		
\usepackage{graphicx}				
\usepackage{amssymb}

\begin{document}

\title{Testing dark matter and geometry sustained circular velocities in the Milky Way with Gaia DR2}

\author{Mariateresa Crosta$^{1}\footnote{corresponding author: mariateresa.crosta@inaf.it}$, Marco Giammaria$^{1,2}$, Mario G. Lattanzi$^{1}$, Eloisa Poggio$^{1}$
\\$^{1}$INAF - OATo, Turin, Italy \\ $^{2}$UNITo, Dep. of Physics, Univ. of Turin, Italy}


 \maketitle

	\begin{abstract}
	Flat rotation curves in disk galaxies represent the main evidence for large amounts of surrounding dark matter. Despite of the difficulty in identifying the dark matter contribution to the total mass density in our Galaxy, stellar kinematics, as tracer of gravitational potential, is the most reliable observable for gauging different matter components. Very recently, the Gaia mission (ESA) has provided kinematics with unprecedented accuracy and consistency across 11 kpc in Galactocentric distances.
	As the Gaia data are analyzed in the framework of Relativistic Astrometry, a consistent general relativistic phase-space picture of the Milky Way (MW) should also be pursued. In this respect, this work aims to test the flatness of the MW rotation curve with a simple general relativistic model suitable to represent the geometry of a disk as a stationary axisymmetric dust metric at a sufficiently large distance from a central body.
	Circular velocities of unprecedented accuracy were derived from the Gaia DR2 data for a carefully selected sample of disk stars, i.e. the angular-momentum supported population of the Galaxy that traces its observed rotation curve. We then fit these velocities to both the classical, i.e. including a dark matter halo, rotation curve model and a relativistic analogue, as derived form the solution of Einstein's equation.  
	The GR-compliant MW rotational curve model results statistically indistinguishable from its state-of-the-art DM analogue. This supports our ansatz that a stationary and axisymmetric galaxy-scale metric could "fill the gap" in a baryons-only Milky Way, suggestive of star orbits dragged along the background geometry.
	We confirmed that geometry is a manifestation of gravity according to the Einstein theory, in particular the weak gravitational effect due to the off-diagonal term of the metric could account for a "DM-like" effect in the observed flatness of the MW rotation curve. In the context of Local Cosmology, our findings are suggestive of a Galaxy phase-space as the exterior gravitational field of a Kerr-like source (inner rotating bulge) without the need of extra-matter.
	
	\end{abstract}

	
	

	\section{Introduction}
	
	Thanks to the Gaia mission \cite{gaia1, gaia2}  the weak gravitational regime is playing a pivotal role in providing a complementary observational perspective for understanding gravity. 
	The few-micro-arcsecond level ($\mu$as) of the Gaia measurements requires a fully general-relativistic analysis of the inverse ray-tracing problem, from the observational data (e.g., stellar images on a digital detector) back to the positions of light-emitting stars (Crosta et al. \cite{crosta17} and references therein). This is because the Gaia-observer is embedded in the ever present and ever changing overlapping weak local gravitational fields of the Solar System. Once the observer is properly defined, null geodesics represent the real physical link through space-time up to the stars. This is the framework of modern Relativistic Astrometry.
	
	By routinely scanning individual sources throughout the whole sky,  Gaia directly measures the kinematics of the stellar component of the Milky Way (MW).
	Gaia's second data release (DR2, Gaia Collaboration \cite{gaia2}) is the first of its deliveries providing parallaxes and annual proper motions, to $\sim$ 100 $\mu$as (for the brighter stars), for about 1.3 billion of the objects surveyed. It also includes Gaia-measured radial velocities (RVs), although for "only" 7 million stars with estimated effective temperatures between 3550 and 6900 $K$ \cite{k18}.
	
	Once a relativistic model for the data reduction is in place, any subsequent scientific exploitation should be consistent with that model. In this respect, our analysis is the first attempt to apply the relativistic kinematics delivered by Gaia to trace the flat Galactic rotation curve at large radii from its center. Such flatness is currently explained as a deviation from the Newtonian velocity profile because of the presence of dark matter \cite{zwicky, verarubin} or of modified gravity \cite{MOND}.
	
	Our quest is pursuing a coherent general relativistic phase-space picture of the MW against which dark matter theories or possible deviations from General Relativity (GR) can be tested. 
	This suggests reconsidering the level of "smallness" and, therefore, "negligibility" usually applied to Galactic dynamics, where the concept of {\it small velocity} is usually used since $v_{Gal}/c \approx 10^{-3}$ for typical galactocentric rotational velocities of disk stars. According to the virial theorem all forms of energy density within the gravitational bound system must not exceed the maximum value of Newtonian potential in it. Regarding the measurements  performed from within the weakly relativistic regime of the Solar System (SS), the lowest order of contribution to the metric (e.g. the term $h_{00}$) is given by the virial theorem as proportional to $(v_{SS}/c)^2 \approx 2 $ milli-arcsecond (mas), requiring the micro-arcsecond ray-tracing modeling for Gaia to include the non-diagonal term $h_{0i} \approx (v_{SS}/c)^3 \approx 0.2 \mu$as. Applying the same reasoning to a conjectural metric for the Galaxy weak gravitational fields away from its center, the non-diagonal contribution is of $\sim$ $(v_{Gal}/c)^3 \approx 100$ $\mu$as, already within the error level of Gaia's DR2. 

	The GR small curvature limit may not coincide with the Newtonian regime, as it is the case of the Lense-Thirring effect \cite{lensthirr}.  
	The situation appears similar to what was needed to explain the advancement of Mercury's perihelion: instead of correcting the dynamics by adding a "dark planet" (Vulcano),  GR cured the anomalous precession by accounting for the weak non-linear gravitational fields overlapping nearby the Sun. Despite it amounts to only 43"/century, because of the small curvature, the effect was  ''strong'' enough to justify a modification of Newtonian theory. 
	On the other hand, in the past it was fruitful to formulate new epistemological interpretations of accurate measurements, presenting new inexplicable features, possibly within the theory underlying them.  
	The {\it aether}, for example, was removed by defining a new kinematics (i.e. the ansatz of special relativity, Einstein \cite{Ein05}) that satisfied the Michelson Morley experiment and Maxwell's equations, instead of adding a new dynamics, i.e. the "extra molecular force" from the Lorentz-FitzGerald contraction effect \cite{LorFitz}, to Newtonian theory. 
	
	Currently, GR is the confirmed standard theory that explains gravity over a range of sixty orders of magnitude. We may certainly assert that the evolution of the MW, and its constituents, is the product of the action of gravity. Nevertheless, only a few exact  solutions of Einstein equation exist, making even the more difficult to detail a metric for the whole Galaxy especially if made of different structures.

	\section{The rotational velocity profile in a stationary axisymmetric spacetime}
	
	Given the premise above, our first attempt to explain the MW rotation curve is to consider a simple relativistic model suitable to represent the Galactic disk as dust in equilibrium at a sufficiently large distance from a (rotating) central body via stationary and axially-symmetric solutions for the disk metric.  In such a space-time there exist two commuting killing vector fields,  $k^{\alpha}$ (time-like) and $m^{\alpha}$ (always zero on the axis of symmetry) and a coordinate system $\{t, \phi, r, z\}$ adapted to the symmetries \cite{fdfclark, stephani}, whose line element for a rotating perfect fluid takes the form: 
		\begin{equation}
		ds^2=  - e^{2U}(d t + A d\phi)^2 + e^{2U} \left( e^{2 \gamma}( dr^2+dz^2) + W d \phi^2  \right),
		\label{metric2}
		\end{equation} 
		$ e^{2 U}, e^{2 \gamma}$ are conformal factors and $U,A, W$ depend only on coordinates $\{r,z\}$. The time coordinate t (time-like far enough from the metric source) spans in the range $[-\infty, +\infty ]$ and $\phi$ is the azimuthal angular coordinate in the range $[0, 2\pi ]$ \cite{fdfclark}. For the general dust solution (\cite{stephani}, eq. (21.50)) we have\footnote{Symbol ( $|$ ) stands for the scalar product relative to the chosen metric.}:
		\begin{equation}
		-e^{2U}=(k|k), \,\, - A e^{2U}  =(k|m), \,\,e^{-2U}W^2 -A^2 e^{2U} =(m|m).
		\end{equation}
		In addition
		\begin{equation}
		m^{\alpha}= \partial^{\alpha}_{\phi}, \,\, k^{\alpha}= \partial^{\alpha}_{t}, \,\,   \partial_{t} g_{ij}=  \partial_{\phi} g_{ij}=0, \,\, g_{\phi a} = g_{t a}=0,
		\label{ortoconditions}
		\end{equation}
		where $a=r,z$.
		Because of the two dimensional Laplace equation one can choose $W=r^2$. 
		
		Let us recall that the dust in GR is defined as a pressureless perfect fluid, i.e. a continuous distribution of matter with stress-energy tensor $T^{\alpha \beta}= \rho u^{\alpha} u^{\beta} $ in geometrized units\footnote{We use greek indices that go from 0 to 3.}, where $\rho$ is the mass density.  A pressure-less fluid may be considered as a very close approximation to a low energy density regime \cite{wald}\footnote{It is worth underlining here that standard $\Lambda$CDM Cosmology is based on the Friedman-Lema\^itre-Robertson-Walker metric, namely a model valid for perfect fluid particles in a homogeneous and isotropic universe.}. Moreover the conservation equation implies geodesicity of the four-velocity and conservation of the mass-energy distribution.  
		Of course, the modeling must avoid the bulge where the axis of symmetry resides (so that $k^{\alpha}$ is time-like globally).
		%
		
		
		For rigidly rotating dust (i.e. shearfree and expansionfree), one can choose $U=0$ and there exist a time-like Killing vector (linear combination of $k^{\alpha}$ and $m^{\alpha}$ with constant coefficient) parallel to the four-velocity of the fluid $u^{\alpha}$, i.e. the co-rotating one chosen by Balasin and Grumiller \cite{bg08}, proportional to $\partial_t^{\alpha}$ (Stephani et al., \cite{stephani}).  Then, by setting $e^{2\gamma}\equiv e^{\nu}$, $N=-A $ and $e^{2U}= 1$, eq. (\ref{metric2}) becomes the line element adopted in eq. (5) of \cite{bg08}. 
		
		In virtue of line element (\ref{metric2}) and conditions (\ref{ortoconditions}), the unit tangent vector field of a general spatially circular orbit can be expressed as
		\begin{equation}
		u^{\alpha} = \Gamma(k^{\alpha} + \beta m^{\alpha}),
		\label{observer-generalcircular}
		\end{equation}
		where $ \beta$ is the constant angular velocity (with respect to infinity) and $ \Gamma$ the normalization factor. Equation~(\ref{observer-generalcircular}) represents a class of observers that includes static ones ($\beta=0$), and can be parametrized either by $\beta$ or equivalently by the linear velocity, say $\mathbf \zeta$, with respect to the ZAMOs ($Z^{\alpha}$, Zero Angular Momentum Observers) as:
		\begin{equation}
		u^{\alpha} = \gamma (e^{\alpha}_{\hat 0} +  \zeta^{\hat \phi} e_{\hat \phi}^{\alpha}),
		\label{obseverzamo}
		\end{equation}
		where $\gamma= -(u|Z)$ is the Lorentz factor, $e^{\alpha}_{\hat 0}$ is the unit normal to the $t$=constant hypersurfaces, and $e_{\hat \phi}^{\alpha} $ the $\phi$ unit direction of the orthonormal frame adapted to the ZAMO. 
		
		ZAMO frames are, indeed, locally non-rotating observers, who have no angular momentum with respect to flat infinity and move on worldlines orthogonal to the hypersurfaces t=constant. The associated tetrad is: $e_{\hat 0}^{\alpha} \equiv Z^{\alpha}$, $e_{\hat \phi}^{\alpha} \equiv 1/\sqrt{g_{ \phi  \phi}} \partial^{\alpha}_{\phi}$, and $e_{\hat a}^{\alpha} \equiv e^{-\nu} \partial^{\alpha}_{a}$.
		
		Then, the line element (\ref{metric2}) can be rewritten in terms of the lapse $M= r/ (r^2-N^2)^{1/2}$ and the shift factor $M^{\phi}= N/(r^2-N^2)$ as 
		\begin{equation}
		ds^2= - M^2 dt^2 + (r^2-N^2) \left(d\phi + M^{\phi} dt \right)^2 + e^{\nu} (dr^2 + dz^2),
		\label{metric2-zamo}
		\end{equation} 
		where $Z^{\alpha}= (1/M) (\partial_t - M^{\phi} \partial_{\phi})$ and the relationship between $\beta$ and $ \zeta^{\hat \phi} $ is given by equating equations (\ref{observer-generalcircular}) and (\ref{obseverzamo})
		\begin{equation}
		\zeta^{\hat \phi} =\frac{\sqrt{g_{\phi \phi}}}{M} ( \beta + M^{\phi}),
		\label{obseverzamostatic}
		\end{equation}
		which, in case of a static observer, reduces to 
		\begin{equation}
		\zeta^{\hat \phi}=
		\frac{N(r,z)}{r}.
		\label{nuzamostatic}
		\end{equation}

The function $N(r,z)$ was solved by BG with the separation anstaz $N(r,z)=F(r) F(z)$ and by assuming reflection symmetry.  
	Their final expression is (eq. 25 in Balasin \& Grumiller \cite{bg08}):
	\begin{equation}
	N(r,z)= V_0 (R-r_0) + \frac{V_0}{2} \sum_{\pm} \left(\sqrt{(z\pm r_0)^2 + r^2} - \sqrt{(z\pm R)^2 + r^2}) \right),
	\label{nrz}
	\end{equation}    
	where the three parameters $V_0 , R, r_0$ were chosen, respectively, as the flat regime velocity, the extension of the MW disk and the bulge radius.  
	Note that $N(r,z)$  was obtained by avoiding values that could prevent a physical solution, such as the localized exotic energy-momentum tensor attributed to \cite{ct07}, or violate the weak energy condition and the assumption of vanishing pressure (see appendix B in Balasin \& Grumiller \cite{bg08} and references therein).  
	Bear in mind also that the Gaia observables are developed with respect to the static observer $u^{\alpha}= (1/{\sqrt{-g_{tt}}}) \partial^{\alpha}_t$ locally at rest relative to the BCRS (in the gravitational fields of the Solar System), which reduces to be $\propto \partial^{\alpha}_t$  far away from it \cite{crosta17}.

Equation (\ref{nuzamostatic}), then, represents the velocity of the co-rotating "dust particle" as measured by an asymptotic observer at rest with respect to the rotation axis and turns out to be proportional to the off-diagonal term $g_{t \phi}$ of the metric~(\ref{metric2}), i.e. the background geometry. Therefore it arises as a relativistic effect due to the gravitational dragging (de Felice and Clarke \cite{fdfclark}). 
	The same applies for a Kerr-like metric\footnote{The vacuum solution of the Einstein field equation for stationary, axisymmetric, asymptotically flat space-time.} where 
	with respect to a suitable tetrad  (de Felice and Clarke \cite{fdfclark})  a static observer has a non-zero angular momentum with respect to infinity, i.e. $(\partial_{\phi}|u) = g_{t \phi}/\sqrt{-g_{tt}}$; on the other hand, ZAMOs have zero azimuthal angular momentum  i.e. $(\partial_{\phi}|Z) = 0$, but a non-zero angular velocity due to the gravitational dragging
.  
	
	In general any particle moving in a metric independent from $t$ and $\phi$ coordinates has two conserved quantities, say, $p_t$ and $p_{\phi}$.
		Consider to drop a particle "radially" from infinity with angular momentum $p_{\phi}=0$; then, $p^{\phi}= g^{\phi t} p_t$ and $ p^t= g^{tt} p_t $. By taking $p^t \propto dt/ d\lambda$ and $ p^{\phi} \propto d\phi/d\lambda$ ($\lambda$ is an affine parameter) it results:
		\begin{equation}
		\frac{p^{\phi}}{p^t}=\frac{g^{\phi t}}{g^{tt}}=\frac{d\phi}{dt},
		\label{angvel}
		\end{equation}
		namely, the particle acquires an angular velocity in the same direction of the rotating gravity source while approaching it \cite{pad}.
		
		The considerations above might suggest a Galactic structure dominated, in the inner part, by a Kerr-like source that, far away from it, turns into a perturbed Schwarzschild-like metric  or a co-rotating "dust" (\ref{metric2}) (for example Lynden-Bell et al. \cite{lynden}).\footnote{According to \cite{stephani}  there exists a one-to-one correspondence between static vacuum solution and (rigidly rotating) dust stationary solution (theorem 21.1).}

	\section{Fit of Gaia-DR2 data to relativistic and classical MW rotation curves}
	
	To study the rotation curve profile of our Galaxy we selected stars tracing the MW disk from the recently released Gaia DR2 archive following the strict criteria illustrated in appendix~\ref{sample}. DR2 directly provided all of the data, i.e. astrometry (parallaxes and annual proper motions) and RVs, necessary for a proper 6-dimensional reconstruction of the phase-space location occupied by each individual star as derived by the same observer. 
	
	At the end of our selection process we are left with a very homogenous sample of 5,277 early type stars and 325 classical type I Cepheides as classified by the Gaia pipelines \cite{clementini}, the largest stellar sample of this kind ever. 
	
	Both spatial and kinematical tests were conducted to ensure that the selected data set fairly traces the MW disk and its kinematics. A close look at the radial and vertical distributions of our sample shows that 99.4 \% (i.e., 5,566) of its stars are  within $4.9 \le r \le 15.8$ kpc (a range of $\sim$ 11 kpc) and below 1 kpc from the galactic plane, that represents the characteristic scale height for the validity of the BG model. 
	
	The quantities extracted from the Gaia DR2 archive are transformed from their natural ICRS reference frame \cite{m18} to its galactocentric cylindrical counterpart, i.e., into the quantities $R$, $\phi$, and $z$ for the galactocentric spatial coordinates and their corresponding velocities $V_R$, $V_\phi$ (i.e., the circular velocity at any galactic longitude), and $V_z$ (further details in appendix~\ref{spakia}).
	
	Therefore, our carefully selected sample of disk stars, the angular-momentum supported population of the Galaxy, allows the reconstruction of the (galactocentric) circular velocities directly from the DR2 data, with which we can trace the observed rotation curve.
	
\begin{figure*}
		\centering
		\includegraphics[width=\textwidth]{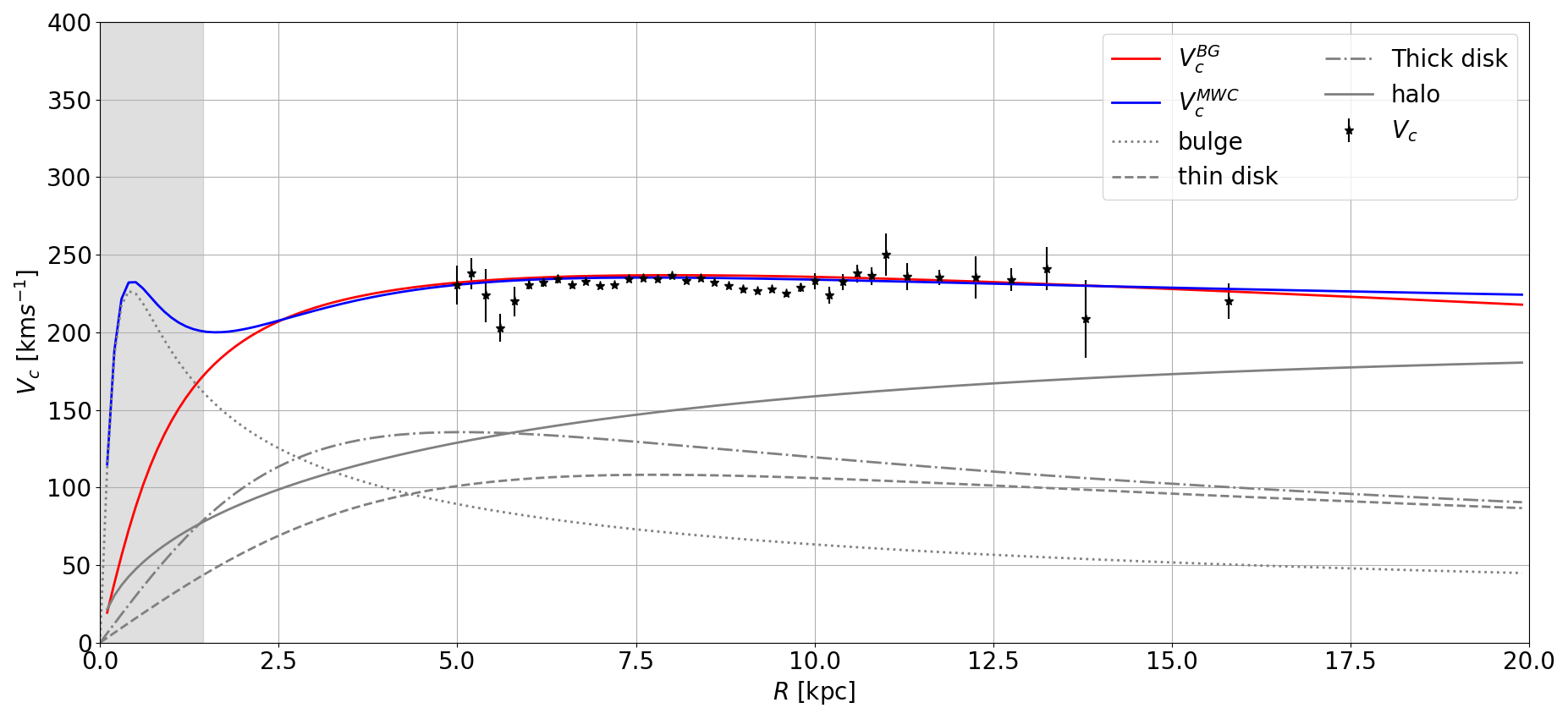}
		\caption{Circular velocity of the MW derived from the DR2 selected disk tracers sample. The black starred symbols represent the medians of the circularized velocity $V_{c}$ derived from equation \ref{eq:vcirc}  with the corresponding error bars that take into account the measurements uncertainties and systematic errors for each radial bin. The red and blue curves show the best fit to the BG and MWC models, respectively.
			The other grey curves represent the kinematical substructures that contribute to the MWC model: the dotted line is the bulge contribution, the dashed and dot-dashed lines that of the thin and thick disk, and the solid line is for the NFW halo.
			The gray vertical band represents twice the value of $r_{in}$ estimated with the BG model.}
		\label{fig:Vrot}
	\end{figure*}

	We bin the data in cylindrical rings as a function of cylindrical coordinate $R$. 
	Table~\ref{table:data} in appendix~\ref{spakia} provides the characteristics of each of the radial bins described in terms of medians and associated RSE's.
	The values for $|z_{median}|$ and the median $V_{\phi}$'s are quite compatible with those expected for a population belonging to the MW young disk and confirm, in turn, the effectiveness of the procedure we adopted for extracting stars from the upper main sequence. 
Nevertheless, as our sample is still "warm", we decided  to use the cylindrical form of the Jeans equation for a axisymmetric disk \cite{j15,bt08}, i.e.,
	
	\begin{equation}
	\frac{\partial (\rho<V_R^2>)}{\partial R} + \frac{\partial (\rho<V_R V_z>)}{\partial z} + \rho \Bigg(\frac{<V_R^2> - <V_\phi^2> + V_c^2}{R}\Bigg) = 0 ,
	\label{eq:jeans}
	\end{equation}
	to circularize our $V_\phi$. This equation links the moments of the velocity distribution $<V_i V_j>$ and the density $\rho$ of a given stellar sample to the circularized velocity $V_c$. The circular(ized) velocity is then
	
	\begin{equation}
	V_c^2 (R) = <V^2_\phi> - <V^2_R>\Bigg(1 + \frac{\partial \ln{\rho}}{\partial \ln{R}} + \frac{\partial \ln{<V_R^2>}}{\partial \ln{R}}\Bigg) ,
	\label{eq:vcirc}
	\end{equation}
where we neglected the contributions of the vertical gradients, and where $<v_i^2>$ represents the averaged squared velocity of the velocity matrix in each bin. Following \cite{eilers}, we utilized the exponential radial density profile $\rho(R) \propto exp(-R/h_r)$ with $h_r$ = 3 kpc. Besides, we notice that in the radial range covered by our data ($\sim$ 5-16 kpc), the radial gradient of $<v_R^2>$  (last term in the parenthesis of equation (\ref{eq:vcirc})) is close to zero. With equation (\ref{eq:vcirc}) providing the measured values of $V_c$'s in each radial bin, the corresponding uncertainties are computed via bootstrapping with 100 re-samples on the the individual values of the azimuthal velocities. The total error bars shown in Figure~\ref{fig:Vrot} and considered in the statistical analysis take also into account possible systematic errors (estimated within 5\%) that the approximations mentioned above could introduce. Finally, it is important to notice that the corrections calculated from equation (\ref{eq:vcirc}) to circularize the $V_\phi$'s are always well below  10\% throughout the radial ranged we have probed.
	
	By setting $z=0$, $r_{in} \equiv r_0$, \textbf{$R \equiv R_{out}$}, $r\equiv R$, and \textbf{$ V_{c}^{BG}(R) \equiv \nu^{ \phi} (r)$} in equations (\ref{nuzamostatic}) and (\ref{nrz}) the relativistic rotational velocity profile writes:
	\begin{equation}
	V_{c}^{BG}(R) = \frac{V_0}{R} \Big(R_{out} - r_{in} + \sqrt{r_{in}^2 + R^2} - \sqrt{R_{out}^2 +R^2} \Big),
	\label{eq:Vbg}
	\end{equation}
	where the unknown parameters $R_{out}, r_{in}$ will result from  fitting to the data of Table~\ref{table:data} after transforming back to regular physical units. In other words, these quantities identify the range for which the 4D spacetime metric used can describe the MW disk as a rotating fluid with cylindrical symmetry. 
	
	We compare this relativistic model with well-studied classical models for the MW (MWC), which we assume to be comprised of a bulge, a double stellar disk (i.e. thin and thick disk) and a Navarro-Frenk-White (NFW) dark matter (DM) halo. Details are in appendix~\ref{grcrotcurve}. We used Markov-Chain Monte-Carlo (MCMC) method to fit to the data; Tables~\ref{table:fit_BG} and~\ref{table:fit_NFW} report  the best fit estimates as the mean of the posteriors and their 95\% confidence interval. For both models, the errors due to the Bayesian analyses are at least one order of magnitude lower than the resulting uncertainties of the parameters. 
	This shows that the analysis is intrinsically consistent and simulation errors are negligible.  
	\begin{table}
		\begin{center}
			\begin{tabular}{c|ccc}
				{\bf BG model }& $\theta$ & $\sigma_\theta^-$ & $\sigma_\theta^+$\\ 
				\hline
				$r_{in} $ [kpc] & 0.9 & 0.4 & 0.4 \\
				$R_{out} $[kpc] & 35.8 & 9.1 & 21.5 \\
				$V_0 $ [km/s] & 296 & 29 & 32 \\
				$e^{-\nu}$ & 0.09 & 0.01 & 0.02 \\
			\end{tabular}
			\caption{$r_{in}$, $R_{out}$ and $V_0$ are the parameters of BG model that correspond to the lower and upper radial limits (i.e. the dimension of the bulge and the {\it Galaxy} radius), and a quantity representing the normalization of the velocity in the flat regime. $e^{-\nu}$ is the conformal factor of line element (\ref{metric2}) at $R_{\odot}$. $\sigma_\theta^-$ and $\sigma_\theta^+$ are the $1-\sigma$ confidential interval of the parameters. 
			}
			\label{table:fit_BG}
		\end{center}
	\end{table}
	
	\begin{table}
		\begin{center}
			\begin{tabular}{c|ccc}
				{\bf MWC model} & $\theta$ & $\sigma_\theta^-$ & $\sigma_\theta^+$\\ 
				\hline
				$M_b [10^{10} {\rm M}_\odot]$ & 0.9 & 0.4 & 0.4 \\
				$M_{td} [10^{10} {\rm M}_\odot]$ & 3.9 & 0.4 & 0.4 \\
				$M_{Td} [10^{10} {\rm M}_\odot]$ & 4.0 & 0.5 & 0.5 \\
				$a_{td} $[kpc] & 5.2 & 0.4 & 0.5 \\
				$a_{Td} $[kpc] & 2.8 & 0.4 & 0.4 \\
				$\rho_0^{halo} [{\rm M}_\odot {\rm pc}^{-3}]$ & 0.010 & 0.003 & 0.004 \\
				$A_h$ [kpc] & 17 & 3 & 4 \\
			\end{tabular}
			\caption{ $M_b$, $M_{td}$, $M_{Td}$, $a_{td}$, $a_{Td}$, $\rho_0^{halo}$ and $A_h $ are the free parameters of the MWC model: the bulge mass, the masses and the scale lengths of the two disks, the halo scale density, and the halo radial scale, respectively. $\sigma_\theta^-$ and $\sigma_\theta^+$ are the $1-\sigma$ confidential intervals of the parameters.}
			\label{table:fit_NFW}
		\end{center}
	\end{table}
	
	In Figure~\ref{fig:Vrot}, the star-like symbols show $V_c$ versus $R$ derived with the Gaia DR2 data in Table~\ref{table:data}. The two estimated velocity profiles are both fairly good representations of the circular velocity. 
	
	We notice that our fit to the MWC model (blue curve) confirms the findings of \cite{i15} that dark matter already contributes above $R > 7$ kpc with a local density estimate of 0,4 GeV/cm$^3$, i.e. $\sim$ 0.01 ${\rm M}_{\odot}/{\rm pc}^{3}$. This compares favorably to our value of  $ 0.009^{+0.007}_{-0.005} {\rm M}_{\odot}/{\rm pc}^{3}$ at $R_{\odot}$, which, in turn, is statistically identical to the very recent value derived in \cite{eilers}.  

	The least constrained parameter in the BG model is the "upper" radial limit, i.e., $R_ {out}$. As already discussed, this was actually expected due to a relatively limited radial coverage of the all--Gaia velocity data we have used. Beside, we obtain an important result on the lower limit parameter  $r_{in}$, which confirms, \emph{a posteriori}, the validity hypothesis of the BG model and the cut at $|z| \le $ 1 kpc we made. In fact, at $R\sim$ 1 kpc it would not be possible to neglect the z-dependence of velocity due to the presence of the MW bulge. 
	
	Finally, our likelihood analysis shows the two models appear almost identically consistent with the data (see appendix~\ref{rvp}).
	
	\section{The local density} 
	
	Once available, the $R_{out}$ and $r_{in}$ parameters can be substituted in the tt-term of Einstein's field equation that in our case results (for its derivation see appendix~\ref{eqdensity}): 
	\begin{equation}
	\rho(R) = e^{-\nu(R)}\frac{1}{8\pi G}\left( \frac{\partial_R N(R)}{R}\right)^2,
	\label{eq:rhobg}
	\end{equation} 
	where $\rho(R)$ is the energy density at $R$ and $e^{-\nu(R)}$ is the conformal metric factor defined in  equation (\ref{metric2-zamo}).
	
	As for the local baryonic matter density, we obtain $\rho(R=R_{\odot}, z=0) \equiv \rho_{\odot}= 0.090 \pm 0.006 M_\odot {\rm pc}^{-3}$ that is in line with independent current estimates (for example, \cite{mc17}, \cite{mk15}, and \cite{mendez}, and references therein).
	
	Our result represents a significant breakthrough with respect to the conclusion of \cite{bg08}.
	Indeed, we do not approximate the density scale factor to some constant value (see section 3.3. of Balasin \& Grumiller \cite{bg08}). Differently, we fit the Gaia data to equation~(\ref{eq:rhobg}) without constraining the density scale factor (which comes from the metric) and compare it with the most recent estimates for the baryonic matter at the Solar position (see also Appendix F).

	\section{Final remarks}
	
	All the observational clues of dark matter point to the existence of a material that: first, it does not absorb or emit light but it exerts and responds only to the gravity force; secondly, it enters the calculation as extra mass required to justify the flat galactic rotational curves. 
	
	By proving our relativistic ansatz on a gravitational dragging effect driving the Galaxy velocity rotation curve,  we suggest that geometry - unseen but perceived as manifestation of gravity according to Einstein's equation - is responsible of the flatness at large Galactic radii. 
	There is nothing new in saying that GR is the standard theory of gravity: we are confirming once more this manifestation by accounting, via the Einstein field equations, for a "DM-like" effect.
	
	It appears that just ''dust'', namely pure matter made only of the non-collisional baryonic mass of the disk, fits the local energy-mass density in accordance with the observations, without further hypothesis according to Occam's razor rule. And the effect vanishes  where no mass-energy density exists. Indeed, our GR rotation curve (Eq. 13 with the constant values given in Table 1) decreases much faster than its MWC counterpart at large Rs, with $V_{c} \sim 100 $ Km/sec at 100 kpc, instead of the ~190 Km/sec value derived from the MWC fit and contributed by the DM halo.
	
	Although these are initial results based on a tailored physical solution of the Einstein field equation,  they show for the first time a possible way out of the dark matter problem in the spirit of the Newtonian {\it Hypotheses non fingo}, suggesting at the same time to push on the use of General Relativity, regardless of how difficult this might be, to detail a more complex Galaxy structure, mostly shaped by the bulky central rotating mass-source.

	\begin{appendix}

		\section{Sample selection}
		\label{sample}
		We selected sources from the recently released Gaia DR2 archive according to the following requirements: (i) availability of the complete astrometric set, and of its corresponding error (covariance) matrix (right ascension $\alpha$ and declination $\delta$, the proper motions $\mu_{\alpha} \dot cos{\delta}$ and $\mu_{\delta}$, and parallax $\varpi$); (ii) availability of the Gaia-measured velocity along the line of sight, $RV$, and its error; (iii) parallaxes good to 20$\%$, i.e., $p/\sigma_{p}\geq$ 5; (iv) availability of a cross-matched entry in the 2MASS catalog \cite{s06}.
		
		Requirements (i) and (ii) are necessary for a proper 6-dimensional reconstruction of the phase-space location occupied by each individual star as derived by the same observer (an important aspect of the relativistic protocol adopted in this article). As for the third criteria, parallaxes to better than 20\% allow us to deal with similar (quasi--gaussian) statistics when transforming them into actual distances, as discussed in \cite{smith-eichhorn96} and references therein. 
		
		Selection criteria (iv) is essential for the actual materialization of the sample of early type stars. For, it provides us with the 2MASS near-infrared magnitudes J, H, and K \cite{s06} that, in combination with the G-band magnitude from the DR2, allow us to build the following photometric filter:

		\[\begin{array}{c}
		(J - H) < 0.14 (G - K) + 0.02  \,\, {\rm and} \,\, (J - K) < 0.23 (G - K)
		\end{array} 
		\]
		
		Following \cite{poggio18}, that needed a stellar sample tracing the MW disk for studying presence and possibly nature of its warp, this filter is then used in combination with their probabilistic method that uses Gaia's astrometry and photometry together to select stars whose colors and absolute magnitudes are consistent with them being upper main sequence stars, including OB stars (see also Re Fiorentin et al. \cite{ReFiorentin19}). 
		On the other hand, as mentioned above, Gaia-measured $RV$'s made the DR2 only when the estimated stellar effective temperatures are between 3550 and 6900 $K$ \cite{k18} for a total of $\sim$ 7.2 million objects. This implies that a large fraction, if not all, of the OB stars initially in the 2MASS cross-matched sample drops out of it because of the $RV$ requirement (ii), leaving us with mainly A, and some F, early type stars. This contingent $RV$-induced bias will be greatly mitigated with the forthcoming Gaia deliveries.

		
\section{Spatial and kinematical analysis}
\label{spakia}
		The Gaia estimated quantities extracted from the DR2 archive are transformed from their natural, i.e. the ICRS \cite{m18}, reference frame to its cylindrical galactocentric counterpart, i.e., into the quantities $R$, $\phi$, and $z$ for the galactocentric position and their corresponding velocities $V_r$, $V_\phi$ 
		and $V_z$. The procedure followed is that described in the \cite{gaiadoc}, and includes proper error propagation thanks to the availability of the correlation matrix (requirement (i)).\\
		For its actual application, we specified the values of the Sun's radial distance in the galactic frame $R_{\odot}$, the Sun's velocity directly in the Galacto-centric reference frame $(U_{\odot}, V_{\odot}, W_{\odot})$, derived from the proper motion of Sgr A* and considered as the Galactic center. In this way, we are independent from the local standard of rest.  
The following values were adopted after reviewing the recent literature: $R_{\odot} = 8.122 \pm 0.031 $kpc \cite{gravity} and $(U_{\odot}, V_{\odot}, W_{\odot}) = (12.9, 245.6, 7.78)$ km/s \cite{dp18}.
		
		We then bin the data in cylindrical rings [$R$-$\Delta R$, $R$+$\Delta R$] as a function of $R$ as described in the caption of Table~\ref{table:data}.
		Finally, we adopt  RSE (from Robust Scatter Estimate) as a robust measure of the dispersion of a distribution. It is defined as $(2\sqrt{2} erf^{-1}(4/5))^{-1} \sim 0.390152$ times the difference between the 90th and 10th percentiles; RSE is the same as standard deviation in the case of a normal distribution.
		
		\begin{table}
			\begin{center}
				\footnotesize{
					\caption{\small{Properties of the binned data for the stellar sample extracted from the Gaia DR2 archive. 
							The data are grouped in cylindrical rings [$R - \Delta R$, $R + \Delta R$] as a function of cylindrical coordinate $r\equiv R$. Each radial bin is centered at the value shown in the second column. The bin size, $\Delta R$, is 0.2 kpc except for the last bins that have been changed to cope with both increasing position errors with distance and the natural decrease in numbers of the Galaxy disk tracers. As robust estimates of the values representing each bin, medians and RSE's are used. 
							The average of the median distances from the plane is $<z_{median}>$=-0,027 in the range between Max($z_{median}$)=0,496 and Min($z_{median}$)=-0,234; moreover, the average value for the vertical dispersion is 0.206 kpc.  As for the azimuthal velocity $V_{\phi}$, the average (across the bins) of the median $V_{\phi}$'s is $\sim$ 224.5 km/sec, while the measured velocity dispersions are always below 41.4 km/s, with a typical (weighted mean) value of 22.1 km/s. }}
					\begin{tabular}{|ccccccc|}
						\hline
						$bin_{size}$ & $R_{mean}$ & $star_{count}$	& $z_{median}$ & $RSE_z$ & $V_{\phi,median}$ & $RSE_{V_\phi}$ \\
						(kpc) & (kpc)  & & (kpc) & (kpc) & (km/s) & (km/s) \\
						\hline
						0.2 &	5.0 &	3 &	-0.234 & 0.072 & 207.3 & 10.9 \\
						&	5.2 &	7 &	-0.077 & 0.114 &	228.4 &	14.9 \\
						&	5.4&	13&	-0.162 & 0.266	& 210.8 & 34.2 \\
						&	5.6&	14&	-0.069 &	0.245 &	204.8 &	21.4\\
						&	5.8 &	30 &	-0.122 &	0.179 &	214.2 &	41.4 \\
						&	6.0 &	40&	-0.112 &	0.171&	225.4 &	37.4\\
						&	6.2 &	71&	-0.125 &	0.217 &	222.9 &	23.3\\
						&	6.4&	102&	-0.124&	0.172&	233.1 &	19.8\\
						&	6.6 &	156 &	-0.078 &	0.181 &	229.5 &	19.3\\
						&	6.8 &	244 &	-0.036 &	0.166 &	228.5 &	19.8\\
						&	7.0&	273&	-0.014&	0.176&	226.0 &	19.2\\
						&	7.2 &	364 &	0.007 &	0.154&	227.3 &	20.2\\
						&	7.4&	392&	0.016&	0.148&	228.4 &	20.1\\
						&	7.6&	428&	0.023&	0.159&	232.6 &	18.7\\
						&	7.8&	366&	0.007&	0.134&	229.9 &	20.4\\
						&	8.0&	368&	0.010&	0.139&	231.7 &	19.7\\
						&	8.2&	342&	-0.010&	0.152&	233.3 &	20.9\\
						&	8.4&	380&	0.009&	0.150&	229.8 &	22.4\\
						&	8.6&	368&	-0.011&	0.143&	228.8 &	23.0\\
						&	8.8&	343&	-0.055&	0.166&	226.8 &	17.2\\
						&	9.0&	296&	-0.054&	0.176&	223.9 &	17.8\\
						&	9.2&	219&	-0.044&	0.189&	222.0 &	18.1\\
						&	9.4&	202&	-0.019&	0.195&	224.2 &	19.6\\
						&	9.6&	155&	-0.039&	0.235&	220.7 &	21.0\\
						&	9.8&	105&	-0.049&	0.240&	222.4 &	20.3\\
						&	10.0&	77&	-0.012&	0.241&	224.2 &	23.1\\
						&	10.2&	51&	0.007&	0.237&	222.6 &	32.9\\
						&	10.4&	27&	-0.067&	0.170&	228.2 &	21.1\\
						&	10.6&	25&	-0.032&	0.290&	227.8 &	22.5\\
						&	10.8&	20&	-0.031&	0.163&	228.4 &	32.3\\
						&	11.0&	13&	-0.103&	0.202&	217.6 &	15.0\\
						&	11.2&	19&	-0.030&	0.250&	233.9 &	27.8\\
						&	11.4&	7&	-0.012&	0.330&	217.2 &	30.9\\
						\hline
						0.5 &	11.75&	18&	0.031&	0.228&	225.5 &	23.5\\
						&	12.25&	20&	0.061&	0.210&	227.8&	21.3\\
						&	12.75&	11&	-0.039&	0.280&	222.6&	18.8\\
						&	13.25&	7&	0.001&	0.287&	230.9&	8.2\\
						\hline
						1  &	13.8&	4&	0.496&	0.386&	217.7&	32.9\\
						1.5  &	15.8&	2&	0.043&	0.420&	219.0&	9.2\\
						\hline
					\end{tabular}
				}
				\label{table:data}
			\end{center}
		\end{table}
		
		
		\section{The fits to relativistic and classical MW rotation curves}
		\label{grcrotcurve}
We assume that the classical MW rotation curves are composed of the following functional components.
For the bulge, we select a Plummer density profile \cite{p17} written as
		
		\begin{equation}
	\rho_{b} (r) = \frac{3 b_b^2 M_b}{4\pi(r^2 + b_b^2)^{5/2}},
		\label{eq:pp}
		\end{equation}
		where, in cylindrical coordinates, the bulge spherical radius is $r = \sqrt{R^2 + z^2}$, with  $b_b = 0.3 kpc$ the Plummer radius \cite{p17} and $M_b$ is the total bulge mass.
		
		 As for the thin and thick MW disks, we use a double-component stellar disk modeled as two Miyamoto-Nagai potentials. This function is also approximated with a double exponential disk as in \cite{mc17} and \cite{kor18}. This is the most general description of a double-component MW disk (Bovy \cite{b15}, Barros et al. \cite{bar16}, \cite{p17}) and we use it in the form
		
		\begin{eqnarray}
		\rho_{d} (R,z)& =& \frac{GM_{d}b^2}{4\pi} \frac{\Big[aR^2 + \Big(a+3\sqrt{z^2+b^2}\Big)\Big(a+\sqrt{z^2+b^2}\Big)^2\Big]} \nonumber  \\
		  &+& {\Big[R^2\Big(a+\sqrt{z^2+b^2}\Big)^2\Big]^{5/2}\Big(z^2+b^2\Big)^{3/2}} ,
	    \label{eq:mn}
		\end{eqnarray}
		where $M_d$ is the total (thin or thick) disk mass, and $a$ and $b$ are scale--length and scale--height. We set $b_{td} = 0.25 kpc$ and $b_{Td} = 0.8 kpc$ as the thin and thick disk scale--heights, respectively.
		
		Finally, we choose a standard NFW model to describe the DM halo (Navarro et al. \cite{nfw96}, McMillan \cite{mc17}, Bovy \cite{b15}) 
		
		\begin{equation}
		\rho_{h} (r) =  \rho_0^{halo} \frac{1}{(r/A_h)(1 + r/A_h)^2}
		\label{eq:nfw}
		\end{equation}
		where $\rho_0^{halo}$ is the DM halo density scale and $A_h$ its (sperical) scale radius. 
		
		The MW total potential can be computed by solving the Poisson equation $\nabla^2 \Phi_{tot} = 4\pi G(\rho_{bulge} + \rho_{td} + \rho_{Td} + \rho_{halo})$; then, the circular velocity follows by solving the differential equation $V_{c}^2 (R) = R \left(d\Phi_{tot}/dR \right)$. 
		We utilized the \emph{GALPY} python package \cite{b15} to calculate each contribution to the classical model, that from now on will be referred to as MWC.
		
		 We fit both the BG and MWC models to the DR2 circular velocities $V_{c}(R_i)$ computed from data in Table \ref{table:data},  and the corresponding uncertainties, utilizing the log likelihood
		\begin{eqnarray}
				\log \mathcal{L}& =& -\frac{1}{2} \sum_{i} \Bigg( \frac{[V_{c}(R_i) - V_{c}^{exp}(R_i|\theta)]^2}{RSE_{V_{c}^i}^2} + \log\Big(RSE_{V_{c}^i}^2\Big)\Bigg) \nonumber \\
		&-& \frac{1}{2}  \Bigg( \frac{[\rho(R_\odot) - \rho^{exp}(R_\odot|\theta)]^2}{\sigma_{\rho_\odot}^2} + \log\Big(\sigma_{\rho_\odot}^2\Big)\Bigg) ,
		\label{eq:like}
		\end{eqnarray}
		where $V_{\phi}^{exp}(R_i|\theta)$ are the expected velocity values evaluated with the two theoretical models at each $R_i$ with a given set of their corresponding parameter vector $\theta$. In this way, the fit takes into account both the uncertainties of the velocity data and the intrinsic non-zero velocity dispersions of the stellar population and possible systematic errors due to the approximations when computing the circular velocity. Moreover, we constrain the local baryonic matter density at the Sun to the most recent estimate of $\rho(R=R_\odot, z=0) = 0.084 \pm 0.012 {\rm M}_\odot {\rm pc}^{-3}$ (see \cite{mk15}). 
		
		For the BG model \cite{bg08}, $\rho^{exp}(R_\odot|\theta)$ is calculated via the 00-term of Einstein equation, while for the MWC model $\rho^{exp}(R_\odot|\theta) = \rho_b(R=R_\odot, z=0) + \rho_{td}(R=R_\odot, z=0) + \rho_{Td}(R=R_\odot, z=0)$ from equations (\ref{eq:pp}) and (\ref{eq:mn}). 
		
	In summary, we have 4 free parameters, $V_0$, $R_{out}$, $r_{in}$ and $e^{-\nu}$, when fitting the BG velocity profile, while we decided for 7 free parameters when dealing with the MWC, i.e. $M_b$, $M_{td}$, $M_{Td}$, $a_{td}$, $a_{Td}$ $\rho_0^{halo}$ and $A_h$.
		
		It is clear that the parameter space is too large to explore with a simple nonlinear fit. We therefore decided to use the Markov Chain Monte Carlo (MCMC) method to determine the unknown parameters and their uncertainties, and  actual computations made use of the MCMC python package PyMC3 \cite{pymc3} with the NUTS algorithms chosen for the step selection. To explore the full pdf we implement the following priors: 
		
\begin{itemize}
\item{\emph{BG model} 
(i) Uniform for $V_0 \in [150, 500] $ km/s; (ii) Uniform for $R\in [10, 300] $ kpc; (iii) Uniform for $r \in [0, 3] $ kpc; (iv) Uniform for $e^{-\nu} \in [0, 1]$;}
			
\item{\emph{MWC model} 
 (i) Normal for $M_b = \mathcal{N}(\mu = 1.067, \sigma = 0.5)\, {10^{10} \rm M}_\odot {\rm pc}^{-3} $; (ii) Normal for $M_{td} = \mathcal{N}(\mu = 3.944, \sigma = 0.5) 10^{10} \, {\rm M}_\odot$; (iii) Normal for $M_{Td} = \mathcal{N}(\mu = 3.944, \sigma = 0.5) 10^{10} \, {\rm M}_\odot$; (iv) Normal for $a_{td} = \mathcal{N}(\mu = 5.3, \sigma = 0.5)$ kpc; (v) Normal for $a_{Td} = \mathcal{N}(\mu = 2.6, \sigma = 0.5)$ kpc; (vi) Normal for $\rho_0^{halo} = \mathcal{N}(\mu = 0.01, \sigma = 0.005)\, {\rm M}_\odot {\rm pc}^{-3}$; (v) Normal for $A_h = \mathcal{N}(\mu = 19.6, \sigma = 4.9) \,{\rm M}_\odot {\rm pc}^{-3}$.}
			
		\end{itemize}
		
In addition, in the MWC model, we fix $b_b = 0.3$  kpc \cite{p17}, as our data can not explore the galactic central region where the bulge dominates. In this way, we eliminate any possible correlations with the free parameters. We stress that for the MWC model we use normal pdf priors so that we could compare our bayesian analysis to the most recent observational estimates (see second item above). On the other hand, we adopted uniform prior distributions for the BG model free parameters in order to avoid any a priori knowledge on quantities never estimated before with MW data.  
		
		 We also fix $b_{td} = 0.25 $ kpc and $b_{td} = 0.8 $ kpc \cite{p17} because in our work we neglect the vertical distribution in the data and consider only binned radial rings as explained above. 
		For the MWC model, the estimated parameters are, within the errors, compatible with literature values \cite{i11, b15, mc17, p17, kor18}. The largest contributions to the $1\sigma$ confidence interval come from the $M_b$ and $A_h$ uncertainties, which are the most difficult to constrain because of the relatively small range covered by the DR2 data. 
		
		The BG solution was first utilized by \cite{dealmeida16} to fit rotational velocity measurements of some external galaxies (i.e. DDO 154, ESO 116-G12, ESO 287-G13,  NGC 2403 2D, NGC 2841 and NGC 3198 1D). The method they proposed consists of converting an observational rotation curve into an effective analogue (called the effective Newtonian velocity profile) that is assumed suitable to be fitted using standard Newtonian gravity procedures. In this way, they state it is possible to apply the Poisson  equation to the relativistic density profile, derived from the BG model, as the non-Newtonian effects are thereby compensated. This is quite a strong working assumption, as some of the terms that derive from GR simply do not exist in the Newtonian weak field approximation. The inadequacy of that assumption might explain the fitting results shown in Table 3 of \cite{dealmeida16}: the parameter \emph{R} (which corresponds to $R_{out}$ in this paper) is in some  cases  unconstrained and somewhat unphysical: $R \sim 10^7$ kpc is beyond what can be considered realistic even for an unusually large and isolated galaxy. The problem  with these results may also be due to the statistical technique used for  the fit, i.e.,  a $\chi^2$ minimization procedure, which we consider insufficient for exploring the parameter space, as explained here.
		
		\section{The goodness of the reconstructed velocity profiles}
		\label{rvp}
		
		Figure 1 shows the two estimated velocity profiles are both good representations of the observed (binned) data. To quantitatively asses this, we compare the two models using the Widely Applicable Information Criterion (WAIC, Watanabe \cite{w10}), which is a fully Bayesian criterion for estimating the out-of-sample expectation.
		
		By definition, lower values of the WAIC indicate a better fit, i.e the WAIC measures the \emph{poorness} of the fit. Our MCMC runs result in the values WAIC = 283.6 and WAIC = 279.8 for the BG and MWC models, respectively. Therefore, for our likelihood analysis the two models appear almost identically consistent with the data.
		
		\section{The Einstein field equations}
		\label{eqdensity}
		Solving Einstein's equation translates into a system of coupled nonlinear partial differential equations, and for that there exist no general method to obtain all of the solutions.
		In fact the Einstein field equation (in geometrized units)
		\begin{equation}
		R_{\alpha \beta} - \frac{1}{2} g_{\alpha \beta} R= 8 \pi T_{\alpha \beta} 
		\end{equation}
		imply finding the expression of the Ricci tensor in function of the metric $ g_{\alpha \beta}$ and the Ricci scalar $R=g_{\alpha \beta} R^{\alpha \beta}$, via the following well known formula:
		\begin{equation}
		R_{\alpha \beta}= \partial_{\lambda} \Gamma^{\lambda}_{\alpha \beta} -  \partial_{\beta} \Gamma^{\lambda}_{\alpha \lambda} + \Gamma^{\lambda}_{\alpha \beta} \Gamma^{\tau}_{\lambda \tau} -\Gamma^{\tau}_{\alpha \lambda} \Gamma^{\tau}_{\beta \tau},
		\end{equation}
		where the Christoffel symbols $ \Gamma^{\lambda}_{\alpha \beta} $ depend on the derivative of the metric according to
		\begin{equation}
		\Gamma^{\lambda}_{\alpha \beta}= \frac{1}{2} g^{\lambda \tau}\left(\partial_{\alpha} g_{\tau \beta} + \partial_{\beta} g_{\tau \alpha}-\partial_{\tau} g_{\alpha \beta}  \right).
		\end{equation}
		Namely, considering line element~(1) and the tensor $T_{\alpha \beta}= \rho g_{\alpha \mu} g_{\beta \tau} u^{\mu} u^{\tau}$ (in virtue of the definition of $T^{\alpha \beta} $ and in the limit of small density ($\rho$), $u^{\alpha}$ results geodetic), one obtains the following expression for the Einstein field equations:
		
		\begin{eqnarray}
		r \partial_z \nu + \partial_r N \partial_z N &=& 0 \\
		2 r \partial_r \nu + (\partial_r  N)^2 - (\partial_z  N)^2 &=& 0 \\
		2r^2 ( \partial_r \partial_r \nu +  \partial_z \partial_z \nu)  + (\partial_r  N)^2 + (\partial_z  N)^2 &=& 0\\
		r ( \partial_r \partial_r N +   \partial_z \partial_z N) -  \partial_r  N &=&0 \\
		(\partial_r  N)^2 + (\partial_z  N)^2  &=& k r^2 \rho e^{\nu} \label{density} 
		\end{eqnarray}
		
		By solving this system of PDE one recovers the functions $N(r,z)$, $\nu(r,z)$ (see section 2.3 and 2.4 in  Balasin \& Grumiller \cite{bg08}), and the mass energy density expression (from eq. ~\ref{density}) we used in our fit to compute the local mass density. 
		
		\section{The derived matter density profile for R $\leq$ 20 kpc}
\begin{figure}
			\includegraphics[width=0.8\textwidth]{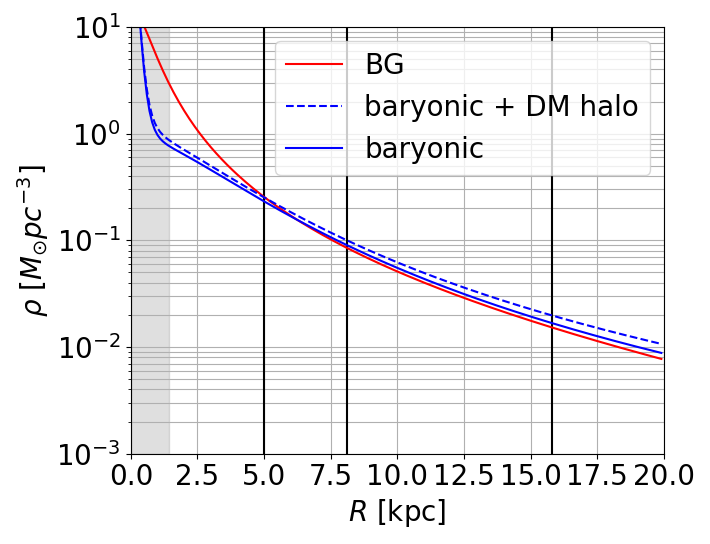}
			\caption{ Density profile of the MW derived from 100 random draws from the posterior distribution of the fit. As in Figure \ref{fig:Vrot}, the red solid line is the BG model, while the blue dashed line represents the total matter contribution for the MWC model (i.e. the sum of the bulge and the two disks as the baryonic counterpart plus the dark matter halo) and the blue solid line the baryonic-matter-only. The black solid lines indicate the range of our data, while the black dashed line is the Sun position in the Galaxy. The grey vertical band represents twice the value of $r_{in}$ estimated with the BG model.}
			\label{fig:dens}
		\end{figure}

		 As described in Appendix \ref{grcrotcurve}, we can now derive the matter density profile for the two models using the best-fit values of their respectively parameters (Table \ref{table:fit_BG} and Table \ref{table:fit_NFW}). We reconstruct 100 random draws from the posterior distribution of the fit and we show the final profiles in Figure \ref{fig:dens}. 
			
		We stress that we use only the observed density values at the Solar position to constraint our likelihood, so what we show are the predicted profiles and not the one of our data. The study of observed density profile is beyond the scope of this work.
		The prediction of the BG model (red line in Figure \ref{fig:dens}) is remarkable and, according to the baryonic-matter-only profile of the MWC model, supports our conclusion that a gravitational dragging effect can produce a flat rotation curve. Moreover, as we use only one data point to set the density normalization, we consider $e^{-\nu}$ as a constant, while in general $\nu = \nu(R,z)$. The future implementation of a independent density profile study will reasonably improves the results and support our thesis.
		
		For $R \leq 5$ kpc the two models have different trends, but this does not affect our analysis because the BG model consider only a single disk structure while the more detailed MWC model takes also into account, e.g., a bulge.
		
	\end{appendix}

	\section*{Acknowledgements}
		This work has made use of data products from: the ESA Gaia mission (gea.esac.esa.int/archive/), funded by national institutions participating in the Gaia Multilateral Agreement; and the Two Micron All Sky Survey (2MASS,www.ipac.caltech.edu/2mass).
		
		We are indebted to the Italian Space Agency (ASI) for their continuing support through contract 2014-025-R.1.2015 to INAF.
		
		Finally, we wish to thank Paola Re Fiorentin, Ronald Drimmel, and Alessandro Spagna for the fruitful discussions on the selection of the stellar sample.

	\end{document}